# Degradation and SEI Evolution in Alloy Anodes Revealed by Correlative Liquid-Cell Electrochemistry and Cryogenic Microscopy

Neil Mulcahy[†], Syeda Ramin Jannat[†], Geri Topore[†], Lukas Worch[†], James O. Douglas[†], Baptiste Gault[†‡], Mary P. Ryan[†], Michele Shelly Conroy[†*]

[†] Department of Materials and London Centre for Nanotechnology, Imperial College London, Exhibition Road, London SW7 2AZ, U.K.

[‡] Max-Planck Institute for Sustainable Materials, Max-Planck Str. 1, 40237 Düsseldorf, Germany

[*] Corresponding author: mconroy@imperial.ac.uk

Understanding solid–liquid interfaces at high spatial and chemical resolution is crucial for advancing electrochemical energy storage technologies, yet this remains a persistent challenge due to the lack of characterisation techniques that can capture dynamic processes and preserve fragile interfacial chemistries. In lithium-ion batteries, interfacial phenomena such as lithium alloying, solid–electrolyte interphase (SEI) formation, and electrode degradation play a decisive role in capacity retention and failure mechanisms but are difficult to observe in their native state due to lithium's high mobility, reactivity, and low atomic number. Here, we use a recently introduced correlative operando characterisation approach that integrates electrochemical liquid-cell transmission electron microscopy (E-LCTEM) with cryogenic atom probe tomography (cryo-APT) to resolve the evolution of a platinum alloy anode at the solid–liquid interface during electrochemical cycling. This correlative, cryo-enabled workflow reveals spatially heterogeneous SEI formation, the presence of lithium-carbonate-rich inner SEI layers, and the retention of elemental lithium within the platinum electrode, most likely trapped along grain boundaries. Additionally, we observe the formation of mossy lithium structures and irreversible lithium loss through dead lithium accumulation. Our results provide direct mechanistic insight into lithium alloying and degradation pathways in alloy-based anodes and establish a generalised platform for probing dynamic electrochemical interfaces with complementary structural and chemical sensitivity. The methodology is broadly applicable to next-generation electrode materials and electrochemical devices where interfacial dynamics dictate performance and stability.

## Introduction

Solid–liquid interfaces in lithium-ion batteries (LIBs) play a critical role in determining electrode performance, stability, and lifetime, yet remain among the most challenging regions to study at high spatial and chemical resolution[1-3]. Processes such as solid–electrolyte interphase (SEI) formation[4, 5] and various degradation mechanisms[6-8] occur dynamically at these buried interfaces and are especially difficult to resolve due to lithium's high mobility, low atomic number, and sensitivity to beam damage[1, 9-11]. A deeper understanding of these nanoscale interfacial transformations is essential for developing next-generation high-capacity anode materials.

Graphite-based anodes have dominated the LIB market for decades owing to their relatively low cost, stable electrochemical profile, long cycle life, and extensive commercial investment[12-15]. Despite these advantages, graphite's specific capacity is limited to 372 mAh/g, restricting the overall energy density of LIBs[14]. With the growing urgency to combat climate change through the adoption of electric vehicles and greener technologies, demand for LIBs with higher energy densities has intensified[5, 16], driving significant research into alternative anode materials such as silicon (Si)[17-19] and lithium (Li) metal[20, 21], which offer greater capacities despite various technical challenges.

Elemental alloying anodes represent a promising alternative to conventional graphite-based anodes[22, 23]. These materials exhibit charge capacities far greater than those of carbon-based anodes, even rivalling the capacities of Si- and Li-metal-based alternatives. In recent years, the use of these alloy materials as foils has gained popularity due to their ease of manufacture, high conductivity, and ability to serve both as a current collector and as a Li host material[22]. A variety of metals, including aluminium (Al), indium (In), lead (Pb), zinc (Zn), gold (Au), and cadmium (Cd), have been proposed for use as alloying anodes[22-25]. While these metals offer promising electrochemical performance, they exhibit distinct electrochemical behaviours, alloying mechanisms, and practical considerations that can significantly influence their suitability and performance as foil anodes in LIBs. Understanding the underlying mechanisms of each type of alloy anode is essential for optimising their performance and ensuring their practical viability in commercial applications.

Although traditionally overlooked as an anode material for commercial applications due to its high cost and low natural abundance, Pt exhibits a high reversible capacity of approximately 600–700 mAh/g — nearly double that of graphite[22, 23]. This high capacity is attributed to Pt's ability to host Li through an asymmetric alloying and dealloying process during cycling, as well as the formation of a SEI layer composed of electrolyte decomposition products[26]. Nonetheless, Pt is routinely used in academic research as a reference electrode[27-29], and Pt foils have found widespread application in liquid cell transmission electron microscopy (LCTEM) as working, reference, and counter electrodes[9, 30-32]. This is largely due to Pt's excellent conductivity, chemical stability, and microfabrication compatibility[33-36]. These properties make Pt particularly well suited for fundamental studies of electrochemical transformations under realistic, liquid-phase conditions.

Li-alloying, SEI formation and degradation of Pt alloy anodes are spatially and temporally dynamic nanoscale processes. The liquid nature of the electrolyte and Li's low atomic number, reactivity and high mobility limits possibilities for quantitative imaging and analysis[1, 9-11]. A deeper understanding of these nanoscale interfacial transformations will not only offer insights into Pts potential as a model anode material but also provides valuable information for broader studies of degradation and SEI formation in advanced LIB systems and energy-storage technology.

Here, we employ a correlative characterisation framework that integrates operando LCTEM with cryogenic atom probe tomography (cryo-APT)[37] to investigate Li alloying, SEI formation, and degradation in a Pt alloy anode. This approach enables real-time imaging of electrochemical transformations in liquid environments, followed by atomic-scale three-dimensional chemical mapping of the preserved electrode–electrolyte interface using cryo-APT, including cryogenic specimen preparation and transfer. Through this multimodal workflow, we uncover spatially heterogeneous SEI growth, Li entrapment, and microstructural cracking—providing mechanistic insight into irreversible capacity loss and establishing a broadly applicable platform for investigating solid–liquid interfacial dynamics in energy storage materials.

## Results and Discussion

### Operando Observation of Mossy Lithium Deposition, SEI Formation, and Pt–Li Alloying Dynamics During Early Charging Cycles

Using a flowing Li electrolyte and a Pt working electrode on a MEMS liquid cell nanochip, we simulated the initial charging stages of a Pt alloy anode by applying four linear voltage sweeps from 0 to –4 V (vs. Pt) at a scan rate of 0.2 V/s, in order to observe Li plating, SEI formation, and the alloying behaviour of Pt with Li. The evolution of the Pt electrode during the second sweep is shown through a series of high-angle annular dark field scanning transmission electron microscopy (HAADF-STEM) images in Figure 1. Movies of all four voltage sweeps are available in Supplementary Movies M1–M4, with corresponding images shown in Supplementary Figure S1. Within this voltage range and at the applied charging rate, both the electrolyte solvent and Li$^+$ ions are electrochemically unstable, leading to competition for electrons during the reaction process and the deposition of their respective reduction products (i.e., SEI components and Li atoms) on the surface of the working electrode[38].

In Figure 1, Li (dark contrast) deposits uniformly around the Pt electrode (bright contrast), appearing at approximately -3.5 V and growing in size with increasing voltage. The Li deposition has a distinct "mossy" morphology[39]. Mossy Li deposits can form due to reaction-limited growth at the electrode, internal stress effects and uneven SEI formation[38, 40, 41]. While relative differences in density allow us to distinguish between the electrochemically active Li deposits, the Pt electrode and the electrolyte, the contrast does not allow easy differentiation between the electrolyte and any formed SEI layer[9]. Furthermore, the limited spatial resolution makes it challenging to determine the extent to which an SEI has formed, its uniformity and its impact on Li deposition on the electrode surface. The application of an intermediate overpotential and a moderate scan rate can result in a SEI growth rate that is insufficient to form a uniform, passivating layer across the working electrode. These non-uniformities in the SEI facilitate localised Li deposition at the electrode surface, which is characterised by the root-growth of mossy Li deposits[38, 42]. The proposed structure of the interface and deposits on the Pt electrode in a given cycle is illustrated in Figure 1 (f).

The growth of mossy Li deposits is visible across all four applied voltage sweeps, as shown in Supplementary Figure S1 (A4, B4, C4, and D4). However, the uniformity of the deposition decreases with continued cycling, largely due to increasing surface roughness of the Pt electrode likely favouring heterogenous nucleation during the subsequent deposition. During charging, the Pt electrode alloys with Li to form metastable Pt–Li compounds, and dealloying occurs upon discharging[22, 26]. This reaction has been reported to be both asymmetric and reversible[26]. In Supplementary Movie M5, the electrode visibly expands and contracts during the voltage sweep, consistent with structural swelling during alloying and shrinking during dealloying. Although the formation of Pt–Li alloys cannot be confirmed from these results alone, their presence has been previously reported in LCTEM studies[9]. With continued alloying–dealloying cycles, electrode roughening becomes more pronounced, with cracking observed in Figure 1(b–e) (highlighted with red circles) and more clearly in Supplementary Figure S1 (D1–D5). Similar microstructural cracking has been reported for both Pt and Au electrodes[9, 43]. The development of such cracking raises questions regarding the full reversibility of cycling in this system, as the accumulation of fatigue, defects, or mechanical stresses is likely facilitating crack nucleation and propagation.

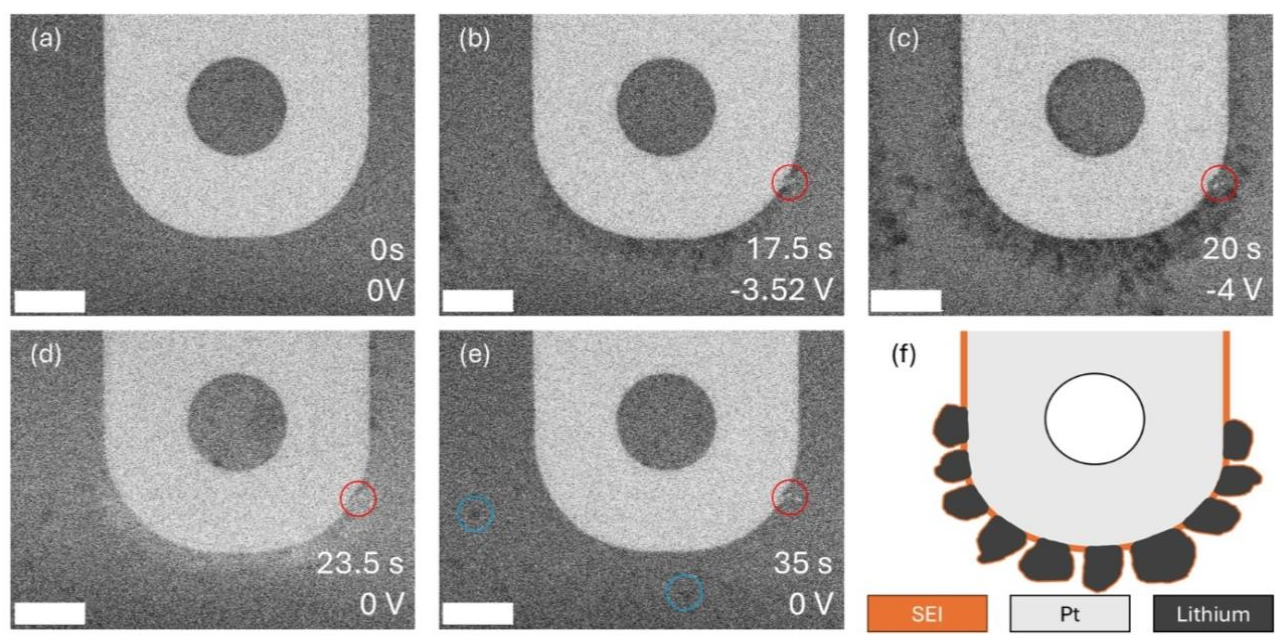

*Fig. 1:* *HAADF STEM images showing the evolution of a Pt working electrode with flowing Li electrolyte when biased from 0 to -4 V. This set of images is displaying the second of four linear voltage sweeps. (a) shows the electrode prior to cycling at 0 V, (b) the electrode with Li mossy deposits appearing approximately around -3.5V with microstructural cracking of the electrode highlighted in red, (c) the system at the maximum applied voltage of -4 V, (d) beginning of stripping process when returned to 0 V and (e) the system following complete stripping. The formation of dead Li is highlighted by blue circles. (f) an illustration of the proposed structure of the deposited species on the Pt electrode surface based on the LCTEM experiments.*

A number of sources of capacity loss have been reported in these systems, most notably through SEI formation and through the formation of dead Li[9, 26, 38]. In Figure 1(e) (blue circles) it can be seen that stripping of the Li mossy deposits is not completely reversible. Upon stripping, some Li becomes electrochemically detached from the electrode, rendering it inactive and effectively "dead" to the system, thereby reducing the overall capacity. The accumulation of dead Li increases with continued cycling, as shown in Supplementary Figure S1 (A6, B6, and C6). Most likely, parts of the Li mossy deposits strip at different rates due to variations in the thickness or presence of SEI layers, with preferential stripping leading to the detachment of some regions[8]. The increased surface roughness of the Pt electrode after repeated cycling favouring heterogenous nucleation is also believed to contribute to the accumulation of electrochemically inactive Li[9]. A 30% reduction in capacity has been reported during the first lithiation cycle for similar systems, largely attributed to irreversible SEI formation[22, 26]. However, upon delithiation, evidence of residual Li trapped within the Pt electrode has also been reported[26]. This trapped Li is associated with irreversible microstructural changes, including minor amorphisation of the electrode[26, 44], while also reducing electrochemical capacity. In the context of the LCTEM results shown here, it cannot be determined with certainty whether remanent Li remains within the system following dealloying.

While operando LCTEM has provided valuable dynamic insights into Li plating, Pt–Li alloy formation, and associated degradation processes, several key questions remain unresolved. These include the presence, composition and structure of a SEI layer, the reversibility of alloying-induced transformations, sources of irreversible capacity from the likes of remanent and dead Li and additionally the origin and progression of microstructural cracking within the Pt electrode. To address these limitations, probing the system compositionally at a higher length scale is essential. Correlative cryo-APT offers the spatial and chemical resolution needed to directly investigate the electrode–electrolyte interface and to resolve nanoscale features that govern electrochemical performance and degradation.

**Freezing and cryo-APT Sample Preparation of the Pt–Li Electrode–Electrolyte Interface**

Following the final cycle of the LCTEM experiment the bottom LC nanochip, containing the electrode of interest covered in a thin layer of electrolyte is frozen using a cold block maintained in liquid nitrogen ($LN_2$)[37]. The frozen interface on the MEMs chip is subsequently transported to the cryo stage of a Plasma Focused Ion Beam/Scanning Electron Microscope (PFIB/SEM) using a Vacuum Cryo Transfer Module (VCTM) via an inert nitrogen glovebox, to prepare APT specimens containing the frozen electrode-electrolyte interface. APT requires the specimen to be needle-shaped with a diameter of approximately 100 nm at the apex[45].

An electrode imaged by SEM in Figure 2 (a) is seen not completely covered by the thin layer of frozen electrolyte. Interestingly grown mossy Li deposits are visible on the edge and on top of the electrode (circled in blue), as well as cracking of the electrode itself, highlighted with red arrows. This provides direct evidence that grown decomposition products survived the LCTEM, freezing and subsequent transfer to the cryo stage of the PFIB/SEM. Using this uncovered electrode as a marker, electrodes covered in electrolyte could be identified, Figure 2 (b), and selectively milled free using the Xe-ion beam. Samples could be lifted out using redeposition welding and attached to a Si post[46-49], Figure 2 (c) and (d). The underside of the sample was filled in with Cr using redeposition to provide mechanical and structural stability to the sample[37, 46, 47]. Using the Xe plasma ion beam the final needle shape containing the electrolyte, Pt electrode and Cr weld could be created with the varying layers readily visible, Figure 2 (e). Each step of the APT sample preparation is described in detail in Supplementary Figure S2, S3 and S4. The needle containing the frozen liquid-solid interface was transferred through ultra-high vacuum and cryogenic temperature via the VCTM into the atom probe, with additional details available in[37].

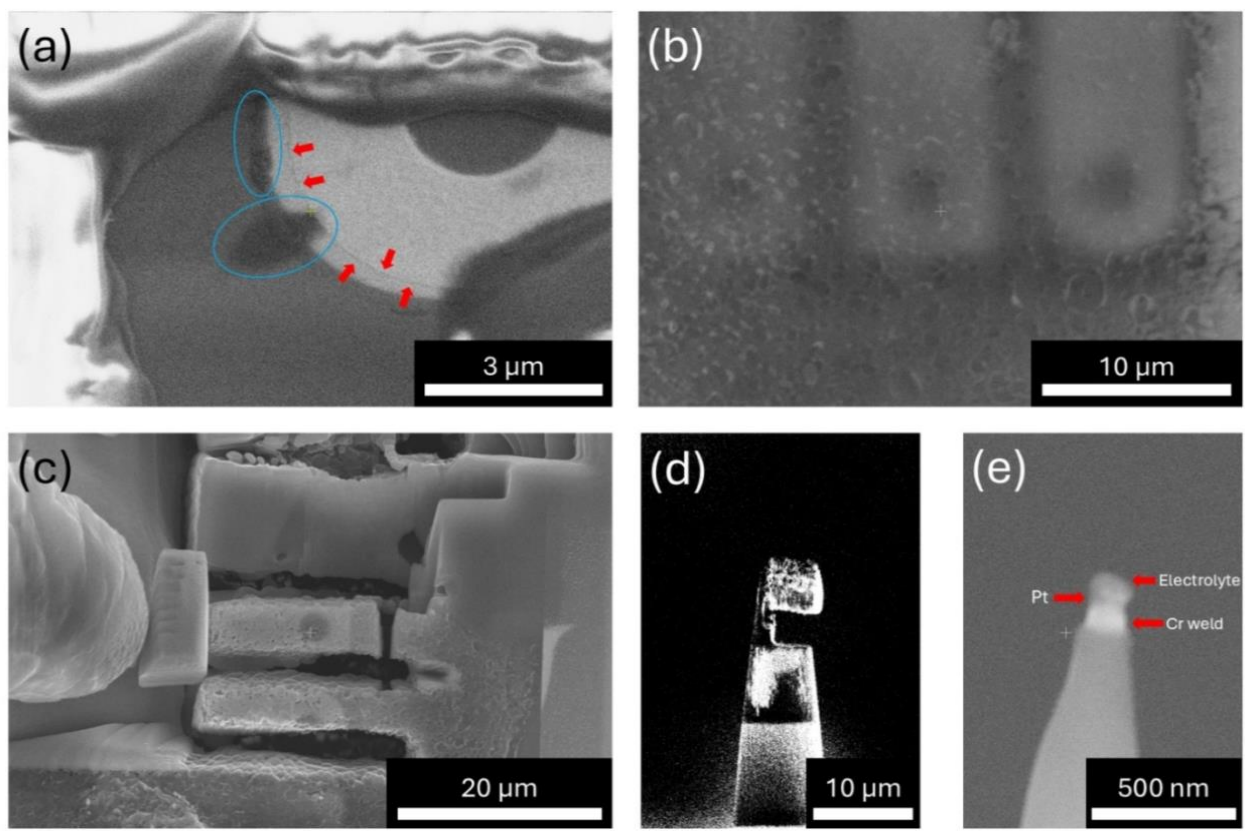

***Fig. 2:*** *SEM & FIB micrographs of (a) an uncovered electrode with visible Li mossy deposits, circled in blue, and cracking along the edge of the Pt electrode (red arrows), (b) a high kV SEM image of electrodes covered under a thin layer of electrolyte, (c) a sample being lifted out using redeposition welding, (d) a sample attached to a post using SEMGlu$^{TM}$, and (e) the final needle sample created with noticeable layers of electrolyte, Pt electrode and Cr weld captured.*

## Bulk chemical analysis via cryo-APT

The APT analysis of the frozen electrolyte–electrode interface produced a mass spectrum and a detector hit map, as shown in Supplementary Figure S5(a) and (b). The electrolyte used in this experiment was $LiPF_6$ in propylene carbonate (PC; $C_4H_6O_5$). For this analysis, detected species containing Li, P, F, O, H, and C ions were classified as originating from the *electrolyte*, while those containing Pt ions were attributed to the *electrode*. Species that contained both Pt and electrolyte-associated elements are discussed in detail below. Hydrogen detection in APT is often affected by background contamination from residual hydrogen present in the instruments analysis chamber. Differentiating between this contamination and hydrogen originating from the specimen is challenging, and current strategies beyond deuteration remain limited[50]. As such, all detected H atoms (including 1,2 Da) in this analysis have been classified as being a part of the *electrolyte*, although this should be interpreted with caution.

Two specific regions of the overall mass spectrum, corresponding to peaks with the highest relative counts, are shown in Supplementary Figures S5(c) and (d). These spectra reveal significant detection of Li and C ions, along with various Pt-containing species, including $Pt^+$, $PtH^+$, $Pt_2P^+$, and $Pt_2C^+$. Ionic identities were attributed to each peak in the mass spectrum and the decomposed species separated into atom type, as shown in Supplementary Figure S5(e).

Notably, detected Li and Pt species account for approximately 80% of the total atomic concentration, indicating that the electrolyte–electrode interface was successfully captured. Several electrolyte-derived species are measured: P (3.49%), F (0.24%), C (7.95%), H (6.70%), and O (1.41%). In addition, trace amounts of Xe (0.18%) were detected suggesting minimal ion beam damage was induced during sample preparation. There is a notable absence of nitrogen-containing species which implies that the plunge-freezing process and subsequent transfer through the nitrogen environment of the glovebox did not introduce detectable contamination or significantly alter the electrolyte composition[37]. Interestingly 0.44% of the detected species involved Cu, which was unexpected in this system. We believe this Cu was contamination from a previous experiment conducted in the LCTEM apparatus using a Cu-based electrolyte. While it is believed that trace amounts of Cu are unlikely to have affected the electrochemical behaviour of the Pt anode in this study, its presence raises concerns regarding trace amounts of contamination in commercial liquid cell systems. All routine cleaning procedures were followed as outlined by the manufacturer. However, this finding suggests that current cleaning protocols when switching electrolytes or solvents may be insufficient. This highlights the need for more stringent decontamination methods and further investigation into sources of trace contaminants within liquid cell systems.

Projections along the -X–Z and Y–Z directions of the 3D reconstruction of the APT data are displayed in Supplementary Figure S6(a) and 6(b), respectively. 2D contour plots displaying the spatial distribution of electrolyte species (H, O, C, P, F, Li), Pt derived species, Cu and Xe in terms relative concentrations along both directions is shown in Supplementary Figure S6(a)(i-iv) and 6(b)(i-iv). The reconstructions reveal two distinct regions: an upper layer rich in electrolyte-derived species, and a lower region composed predominantly of Pt and Li.

## Cryogenic compositional mapping of the SEI

Focusing first on the upper, electrolyte-rich region, decomposed 1D concentration profiles along both the X and Y directions were extracted and are shown in Supplementary Figure S7. These profiles show that over 70% Li concentration is sustained across much of the central region of interest (ROI), with increased concentrations of other electrolyte species (e.g., C, O, P, F, H) observed near the edges. This suggests that the central area primarily contains pure Li or Li-rich species. Figure 3 provides a breakdown of the Li-containing species and other electrolyte species detected in this region, displayed through 2D contour plots of their spatial distribution in terms of relative concentration in the -X-Z orientation. Notably elemental Li is concentrated within the centre of the ROI, while species including $Li_xC$ and $Li_xC_xH_x$ are distributed around these central zones, with large concentrations of elemental C also present, Figure 3(a-d). Furthermore, as seen in Figure 3(e-g), most other electrolyte species, including H, O, P, F, and minor quantities of C and Li, are located even further outside these regions. The trace detection of these other elements within the Li-rich central regions is likely a result of trajectory overlaps during the APT analysis.

Previous studies suggest that the detection of Li–C species in APT may indicate the presence of lithium carbonate ($Li_2CO_3$)[51], which is commonly found in SEI layers. Quantification of all species from the carbonate remains challenging. Complex ions such as $CO_3^+$ may fragment during the flight, following field evaporation, making it difficult to accurately detect all components. Some reports have suggested that neutral molecules like $O_2$ or $CO_2$ can form from dissociation of an initially charged particle, and be lost from the analysis[52]. Additionally, high evaporation rates of lighter elements like Li and C can cause detector saturation and associated losses[51, 53].

Studies focused on compositionally analysing the SEI layer formed on Pt working electrodes at high resolutions remains limited, particularly while maintaining the sample in its native state. Kim et al.[26] reported ex-situ XPS measurements of SEI formation on Pt nanoparticles and showed the reversible formation of LiF and $Li_2CO_3$ as the inner layer of the SEI layer. In the context of these results the APT needle is believed to capturing a portion of the inner part of the SEI layer formed of $Li_2CO_3$.

A simplified diagram of the electrolyte-derived species in this top ROI is displayed in Figure 3(h). Spatially the right-hand side of the top ROI is dominated by carbon and $Li_xC_x$ species, with the interface comprising of both elemental Li and C in contact with the electrode, indicative of the lithium carbonate species comprising the inner layer of the SEI. The central region in contrast, contains a high concentration of lithium with minimal carbon or carbonate species. Although some uncertainty in the reconstruction may arise from lithium migration effects, this area displays characteristic "plumes" of lithium[54]. These features have previously been associated with field-induced adatom gas formation of pure lithium metal[54]. The detection of pure lithium in this region likely reflects the composition of a mossy lithium deposit

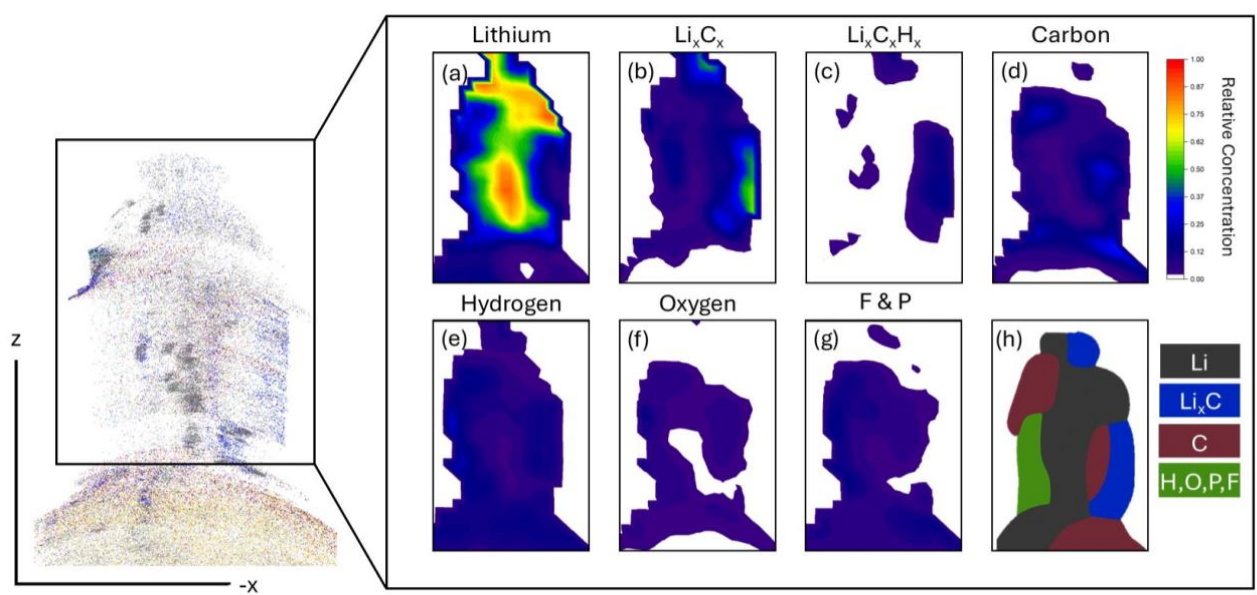

*Fig. 3:* 2D contour plots detailing the spatial distribution of various electrolyte derived species in terms of relative concentrations across the top ROI in the -X-Z orientation. Species include (a) elemental Li, (b) Li-C species, (c) Li-C-H species, (d) elemental Carbon, (e) decomposed H, (f) decomposed O and (g) decomposed F and P. (h) shows a simplified diagram illustrating the relative distribution of the various species within the ROI. Relative concentrations are described in terms of atomic % per unit area.

**Cryogenic compositional mapping of the Pt-electrode**

Based on Supplementary Figure S6, the bottom half of the reconstruction is dominated by Pt, consistent with the Pt electrode used for this experiment. Interestingly, Li is also detected within this region, suggesting Li exists within the electrode structure. In Figure 4 2D contour plots are used to display the spatial distribution in terms of relative concentration in both -X-Z and Y-Z orientations of all Pt containing species, including elemental Pt, $Pt_2P$, PtH and $PtC_2$, as well as elemental Li. There is a notable absence of well-defined Pt–Li compounds, with only elemental Li and Pt observed, rather than alloying species such as LiPt or $Li_2Pt$. This would be expected based on the anode being discharged prior to freezing in $LN_2$ and reflects the reversible nature of the alloying–dealloying process.

The presence of Li within the electrode, as can be seen in Figure 4 (a)(vi) and (b)(vi), is most likely due to remnant Li becoming trapped at grain boundaries or other structural features during cycling[26]. Supporting this, Figure 5 presents 2D contour plots of Li ion concentration at increasing depths within the electrode. The relatively consistent spatial distribution of Li across these slices suggests that the trapped Li is not randomly dispersed but rather localised at specific features, potentially grain boundaries or other structural defects.

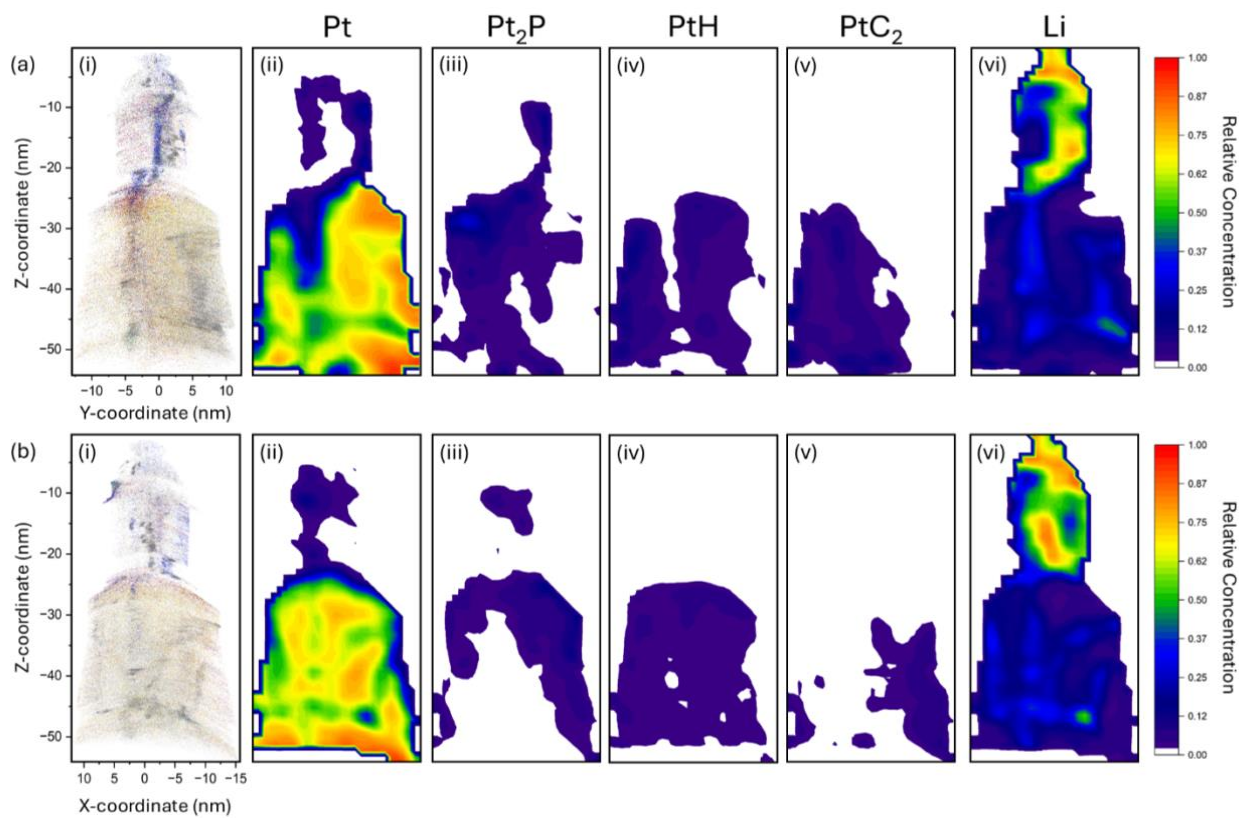

***Fig. 4:*** *2D contour plots detailing the spatial distribution of Pt-derived species and Li in terms of relative concentrations in both (a) -X-Z and (b) Y-Z orientations. Species of note include (ii) elemental Pt, (iii) $Pt_2P$, (iv) PtH, (v) $PtC_2$ and (vi) decomposed Li. Relative concentrations are described in terms of atomic % per unit area.*

Previous reports have attributed irreversible capacity losses of approximately 30% during the first lithiation cycle due to the retention of inactive Li within the electrode structure[26]. Here we provide direct evidence of Li entrapment along grain boundaries, offering a mechanistic explanation for this capacity loss. The remanent Li in this system is most likely becoming confined due to subtle microstructural changes induced during alloying and dealloying. A comparison between a cycled and uncycled Pt electrode in Supplementary Figure S8 further highlights cycling-induced structural evolution. Reports of lithiation occurring almost exclusively along grain boundaries have been shown to occur in other metals such as Silver (Ag)[55], suggesting that a similar mechanism is active in the Pt system shown here. Additionally, the accumulation of Li in confined regions may impose localised strain, potentially contributing to the cracking observed in Figure 1. Repeated cycling and progressive Li entrapment may exacerbate this mechanical degradation over time, consistent with the increased cracking seen across successive voltage sweeps in Supplementary Figure S1.

Various other Pt-electrolyte derived species have been captured, notably $Pt_2P$, PtH and $PtC_2$, Figure 4(a)(iii-v) and (b)(iii-v). These species are predominantly located at the interface between the electrolyte and the electrode, as expected, confirming that the two regions are indeed in contact. Notably, small but distinct concentrations of Pt and Pt-derived species were identified within the electrolyte region itself, as seen most clearly in Figure 4.

In this instance polycrystalline Pt was used as the working electrode, which is common in the type of MEMs nanochip electrochemical cell used. This is primarily due to Pts stability and inertness within various different electrochemical environments. However, it has been shown that under certain aggressive conditions (high overpotentials, low pH environments, certain electrolytes etc.) Pt dissolution can occur and can even lead to Pt redepositing elsewhere within the system[56]. Furthermore, Pt dissolution has been reported to increase significantly at lower scan rates, which are commonly employed in nanoscale LCTEM electrochemical experiments[9, 57]. In this study, the anodic dissolution of the Pt electrode is directly evidenced by the detection of Pt species within the electrolyte region in the APT reconstruction.

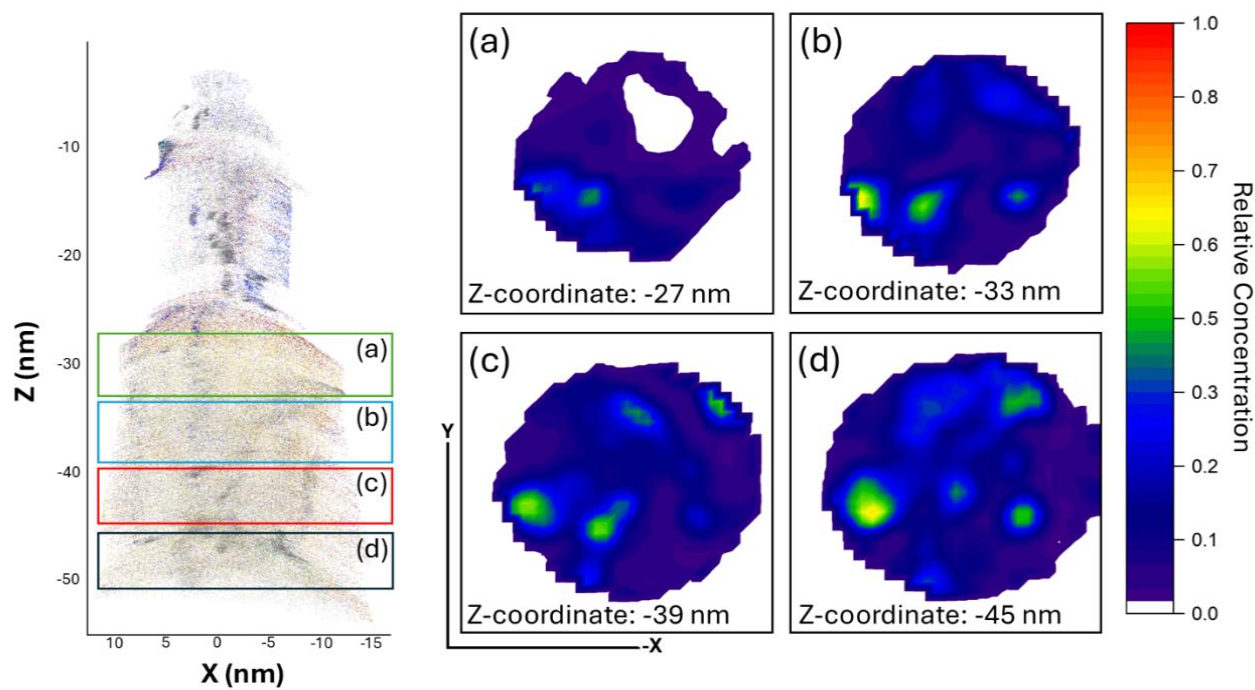

*Fig. 5:* *2D contour plots displaying the spatial distribution of Li at varying Z depths within the electrode in terms of relative concentration. 0 nm corresponds to the top of the reconstruction and slices are displayed at relative depths of (a) -27 nm, (b) 33 nm, (c) -39 nm and (d) -45 nm. Relative concentrations are described in terms of atomic % per unit area.*

**Cryo-APT: perspective on accuracy and precision of the results**

Li is notoriously difficult to quantitively characterise owing to its high mobility, low atomic weight and beam sensitive nature[1, 9-11]. While APT has emerged as an exceptional candidate for analysing Li based systems owing to its high chemical sensitivity, sub-nanometre spatial resolution and cryogenic capabilities, challenges still remain in reliably analysing Li-containing materials[2, 53]. One major issue is in-situ delithiation[2, 58, 59], which can occur when Li-containing samples are exposed to the intense electric fields required to induce field evaporation during APT analysis. Li deintercalated from the structure migrates on the specimen's surface prior to field evaporation, thereby compromising the accuracy and reliability of the analysis[54, 59].

To mitigate delithiation effects, various strategies have been proposed and implemented, such as varying sample preparation conditions[2], transfer procedures or applying shielding via in-situ sputtering to suppress delithiation effects[46, 53, 60]. However, many of these methods were not applicable in the present study due to the liquid nature of the electrolyte and to avoid any risk of damaging sensitive SEI components or metallic Li potentially present in the electrolyte region at the sample's surface. In addition, in a similar frozen electrolyte analysed by APT, Kim et al. did not report signs of delithiation[7].

To act as a direct comparison, an APT needle of frozen uncycled electrolyte was prepared and analysed. The created needle is shown in Supplementary Figure S9(a), with the corresponding APT reconstruction presented in Supplementary Figure S9(b). A comparison between the ranged decomposed species and mass spectrum regions representing surface compositions from both the cycled electrode (Supplementary Figure S9(c)) and the uncycled electrolyte is shown in Supplementary Figures S9(d), (e), and (f). Several notable differences are evident between the two datasets. In particular, significantly higher concentrations of H, O, F, and P were detected in the uncycled electrolyte, consistent with its known formulation. In contrast, the cycled system exhibited a much higher Li content, accounting for 62.00% of the detected species, compared to just 1.82% in the uncycled case. Additionally, lithium-carbon compounds ($Li_xC$), notably $Li_3C$, were identified in the cycled sample but were absent from the uncycled electrolyte, highlighting a clear distinction in the nature of species present before and after electrochemical cycling.

## CONCLUSIONS

In this study, we have demonstrated the power of a correlative operando LCTEM and cryo-APT workflow to unravel the complex nanoscale processes governing Li plating, SEI formation, and degradation in Pt alloy anodes. Our operando LCTEM observations revealed the dynamic evolution of mossy Li deposits and SEI growth, as well as the onset of microstructural cracking and irreversible capacity loss through the formation of dead Li. Cryo-APT enabled near-atomic-scale compositional mapping of the electrode–electrolyte interface, providing direct evidence for the spatial distribution and chemical nature of SEI components and mossy Li deposits at the interface, as well as Li entrapment within the electrode structure itself. Together, these complementary techniques have elucidated key degradation mechanisms such as non-uniform SEI formation leading to mossy Li growth, dead Li accumulation and minor electrode roughing and microstructural changes leading to Li entrapment, which all contribute to irreversible capacity loss in Pt alloy anodes. The insights gained offer a robust framework for understanding alloy-based anode behaviour and highlight the critical need for advanced characterisation methods to guide the design of durable, high-performance electrodes for next-generation LIBs. Our approach is broadly applicable to other emerging anode materials, paving the way for rational engineering of interfaces and microstructures to enhance battery performance and longevity.

## MATERIALS AND METHODS

### Liquid cell Scanning Transmission Electron Microscopy

(S)TEM measurements were completed using a Thermo Fisher Scientific (Waltham, Massachusetts, United states) Spectra 300 (S)TEM. This is a probe corrected instrument fitted with an ultra-high-resolution X-FEG Ulti-monochromator. STEM measurements were taken at 300 kV accelerating voltage. The convergence angle was 31.3 mrad, and the collection angle for ADF images was 49–200 mrad using the HAADF detector. All STEM images were processed using Thermo Fisher Scientific Velox software. The measured screen current used was 1pA. This equated to an electron dose of 84 e/Å$^2$. Prior to any biasing, the system was imaged under liquid flow conditions for ten minutes to confirm that the electron beam had no observable impact on the system. This ensured that no beam-induced effects were present and that the phenomena observed during subsequent experiments was not influenced by electron beam irradiation. High angle annular dark field (HAADF) was selected as the imaging mode. It has previously been shown that HAADF-STEM imaging allows for distinct quantification between deposited Li metal, SEI components, and the electrolyte due to their relative differences in density[9].

LCTEM experiments were performed using the Stream system supplied by DENSsolutions B.V. (Delft, The Netherlands). This included the Stream in situ liquid TEM holder, the pressure based liquid supply system (LSS), and the liquid biasing nano cells. Each nanocell is composed of a top and bottom Si wafer chip. The bottom chip is essentially a microdevice containing a working, reference and counter electrode made of Pt. The working electrodes are deposited on a 50 nm SiN$_x$ electron transparent membrane window with dimensions of approximately 20 μm x 200 μm[30]. Combined with an identical electron transparent membrane on the top chip, this allows for viewing of the working electrode while liquid is flowing and electrical biasing is occurring within the TEM, essentially creating a "nanobattery" within the microscope. Prior to insertion into the microscope the bottom nanochip was plasma cleaned in Ar-O for approximately two minutes. This was done to change the properties of the bottom nanochip from being hydrophobic to hydrophilic. This ensured there would be a more uniformly distributed layer of electrolyte over ROI on the nanochip during freezing. This process does lead increased bubble formation during biasing and imaging. However this could be counteracted using a high flow rate and a lower electron dose.

Operando LCTEM electrochemistry experiments were performed using a commercially acquired lithium electrolyte, in this instance $LiPF_6$ in propylene carbonate (PC), supplied by Merck Life Science UK Ltd (Dorset, United Kingdom). To achieve a base flow rate of ≈ 8 μL/min, LSS inlet and outlet pressures were set at 2000 and -950 mbar respectively. The flow rate and inlet/outlet pressures were controlled using Impulse, a commercially available software supplied by DENSsolutions B.V. (Delft, The Netherlands). Linear sweeps and cyclic voltammetry (C.V.) measurements were performed using PS trace 5.6, with a connection running directly from the holder to the LSS controlling the electrochemical biasing. Four linear voltage sweeps were performed. The first three from 0 to -4 V with a scan rate of 0.2 V/s and a step of 0.01 V, while the final sweep was held at -4V for an additional 25s.

**Plasma Focused Ion Beam/Scanning Electron Microscope with Cryogenic capabilities**

A Helios Hydra CX (5CX) plasma FIB from Thermo Fisher Scientific (Waltham, Massachusetts, United states) fitted with an Aquilos cryo-stage and an Easylift tungsten cryo-micromanipulator was used for all FIB/SEM work shown. For the cryogenic work shown here, the stage and manipulator base temperature were set at approximately 133 K and 110 K. This was achieved through the circulation of gaseous nitrogen passing through a heat exchanger within a dewar of $LN_2$, with the flow of gaseous nitrogen being maintained at 180 mg/s. The temperature of the stage and micromanipulator can be directly controlled through the use of a temperature control unit, supplied by LakeShore Cryotronics Inc (Westerville, Ohio, United States), Model 335 cryogenic temperature controller, and heaters within the stage. The FIB column was set at 52° to the electron column, while Xenon plasma was used for all FIB work shown. A "Dual-puck" holder stage baseplate supplied by Oxford Atomic (Oxford, United Kingdom) was fitted to the cryo-stage, which allowed two industry standard cryo pucks, supplied by CAMECA Inc. (Gennevilliers, France), to be inserted to the cryo stage at once. One puck would contain the frozen liquid cell nanochip, while the other would contain a preprepared Si microarray coupon, as detailed in [37, 48]. The system is equipped with a Ferroloader docking station supplied by Ferrovac GMBH (Zürich, Switzerland). This allows samples to be inserted directly to the cryo stage of the PFIB/SEM from a precooled Vacuum Cryo Transfer Module (VCTM), supplied by Ferrovac GmbH (Zürich, Switzerland), while being maintained under vacuum and at cryogenic temperatures.

**Cryogenic Atom Probe Tomography**

A CAMECA Inc. (Madison, WI, USA) Local Electrode Atom Probe 5000 XR was used for all atom probe analysis shown. This instrument is equipped with a reflectron system and a Ferroloader docking system. Samples from a precooled VCTM could be inserted directly into the analysis chamber of the instrument through the use of a "piggyback" puck, while being maintained under vacuum and at constant cryogenic temperatures. The sample was run using laser pulsing analysis (50 pJ, 80-140 kHz, 1 ion per 100 pulses on average, 50k base temperature). Atom probe data analysis and 3D reconstructions and were completed using AP suite 6.3, a commercially available software from CAMECA Inc. (Madison, WI, USA). Due to field-induced aberrations and migration effects, only a subset of the full reconstruction was analysed. Specifically, regions exhibiting consistent spatial resolution and minimal distortion were selected to ensure data reliability and interpretability. This approach ensured the influence of field evaporation artifacts on compositional and structural analyses was minimised.

**Vacuum Cryo Transfer Module (VCTM) and inert glovebox**

A VCTM, supplied by Ferrovac GmbH (Zürich, Switzerland), was used to transfer samples under vacuum/inert conditions and constant cryogenic temperatures between instruments possessing Ferroloader docking stations. The module contains a small ion pump with a non-evaporable getter cartridge, enabling it to sustain ultra-high vacuum levels down to $10^{-10}$ mbar. Cryogenic temperatures within the module are maintained using a $LN_2$-filled dewar. The module is compatible with industry-standard pucks and cryo pucks provided by CAMECA Inc. (Gennevilliers, France). A 500 mm wobblestick with a PEEK-insulated manipulator allows for precise puck handling within the system[61].

All cryogenic sample handling and preparation took place within an inert atmosphere glovebox supplied by Sylatech Ltd. (York, United Kingdom). The glovebox maintains a nitrogen environment with oxygen and humidity levels typically below 5 ppm during operation. $LN_2$ can be introduced into an internal bath from an external high-pressure $LN_2$ dewar, Apollo 50, Cryotherm Inc., (Kirchen (Sieg), Germany). Samples may be plunge frozen directly within the glovebox and subsequently transferred to the VCTM using a combination of the $LN_2$ bath and a cooled elevator system located in a UHV loadlock chamber. This chamber is interfaced with a Ferroloader docking station. Frozen samples are retrieved from the elevator using a wobblestick within the VCTM, ensuring that they remain under vacuum and at cryogenic temperatures throughout the transfer process.

## Acknowledgements

N.M., M.P.R., M.S.C. acknowledge funding from Engineering and Physical Sciences Research Council (EPSRC) and Shell for funding through the InFUSE Prosperity Partnership (EP/V038044/1). This work was made possible by the EPSRC Cryo-Enabled Multi-microscopy for Nanoscale Analysis in the Engineering and Physical Sciences EP/V007661/1. M.S.C. acknowledges funding from Royal Society Tata University Research Fellowship (URF\R1\201318) and Royal Society Enhancement Award RF\ERE\210200EM1. L.W. and G.T. EPSRC Centre for Doctoral Training in the Advanced Characterisation of Materials (CDT-ACM)(EP/S023259/1) for funding their Ph.D. studentships and G.T. acknowledges Cameca Ltd. for co-funding their PhD. SRJ thanks PhD funding from the Faraday Institution, under the grant EP/S514901/1. B.G. acknowledges financial support from the ERC-CoG-SHINE-771602. M.P.R acknowledges support from the Armourers and Brasiers Company.

# Degradation and SEI Evolution in Alloy Anodes Revealed by Correlative Liquid-Cell Electrochemistry and Cryogenic Microscopy

Neil Mulcahy[†], Syeda Ramin Jannat[†], Geri Topore[†], Lukas Worch[†], James O. Douglas[†], Baptiste Gault[†‡], Mary P. Ryan[†], Michele Shelly Conroy[†]

† Department of Materials and London Centre for Nanotechnology, Imperial College London, Exhibition Road, London SW7 2AZ, U.K.

‡Max-Planck Institut für Eisenforschung GmbH, Düsseldorf 40237, Germany

*Corresponding author: mconroy@imperial.ac.uk

**Supplementary Movie M1:** Video showing the evolution of the Pt working electrode during the first voltage sweep

**Supplementary Movie M2:** Video showing the evolution of the Pt working electrode during the second voltage sweep

**Supplementary Movie M3:** Video showing the evolution of the Pt working electrode during the third voltage sweep

**Supplementary Movie M4:** Video showing the evolution of the Pt working electrode during the fourth voltage sweep

**Supplementary Movie M5:** Video of the second voltage sweep at a high frame and repeated to highlight the swelling and contraction of the Pt working electrode during alloying and dealloying

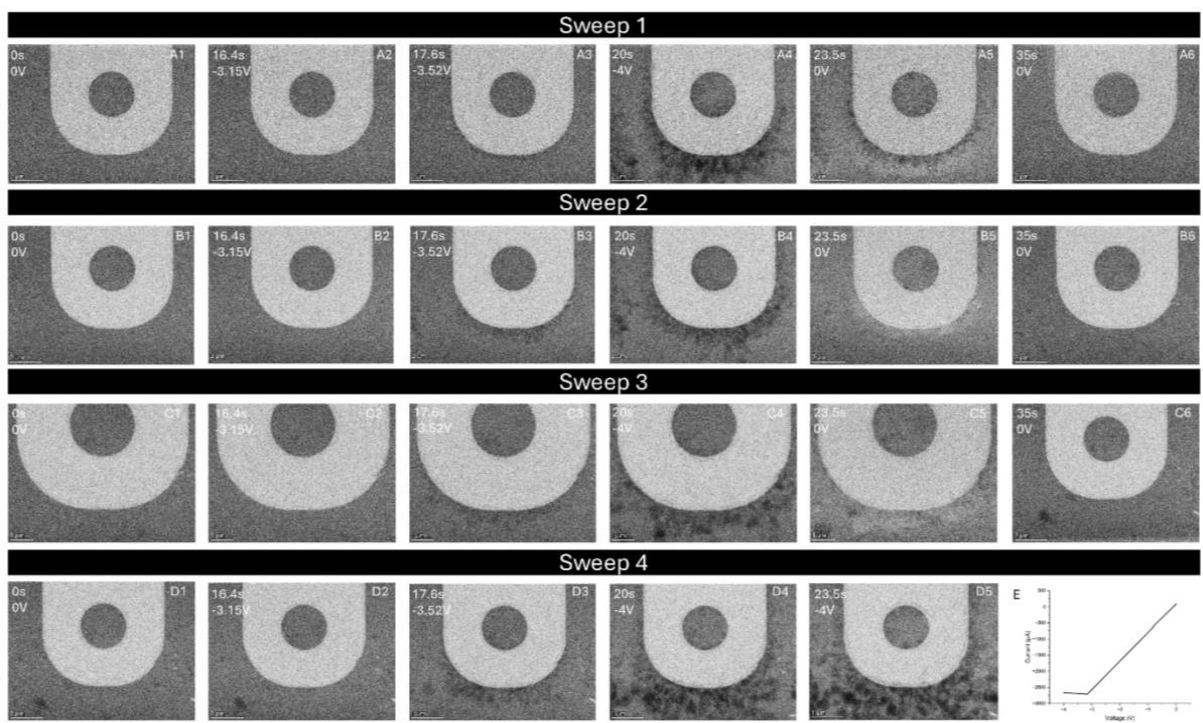

**Fig. S1:** HAADF STEM images highlighting the evolution of the Pt working electrode during each voltage, the light contrast representing the Pt electrode and the dark contrast showing Li. A1-D1 shows the electrode prior to the beginning of each sweep at 0 V. A2-D2 shows the electrode at the point the current reached a steady state at -3.15 V. A3-D3 shows the beginning of Li mossy growth around -3.5 V. A4-D4 shows the point at which the max voltage is applied, -4V, and the point the mossy Li growth is at its greatest extent. A5-C5 shows the point at which the voltage was returned to 0 V and stripping of the Li deposits began, and A6-C6 show the point the stripping has completed for the first three sweeps. For Sweep 4 the electrode was held at -4V for an additional 25 seconds before being removed from the TEM, and this extended Li mossy growth can be seen in D5. An example of an applied voltage sweep can be seen E.

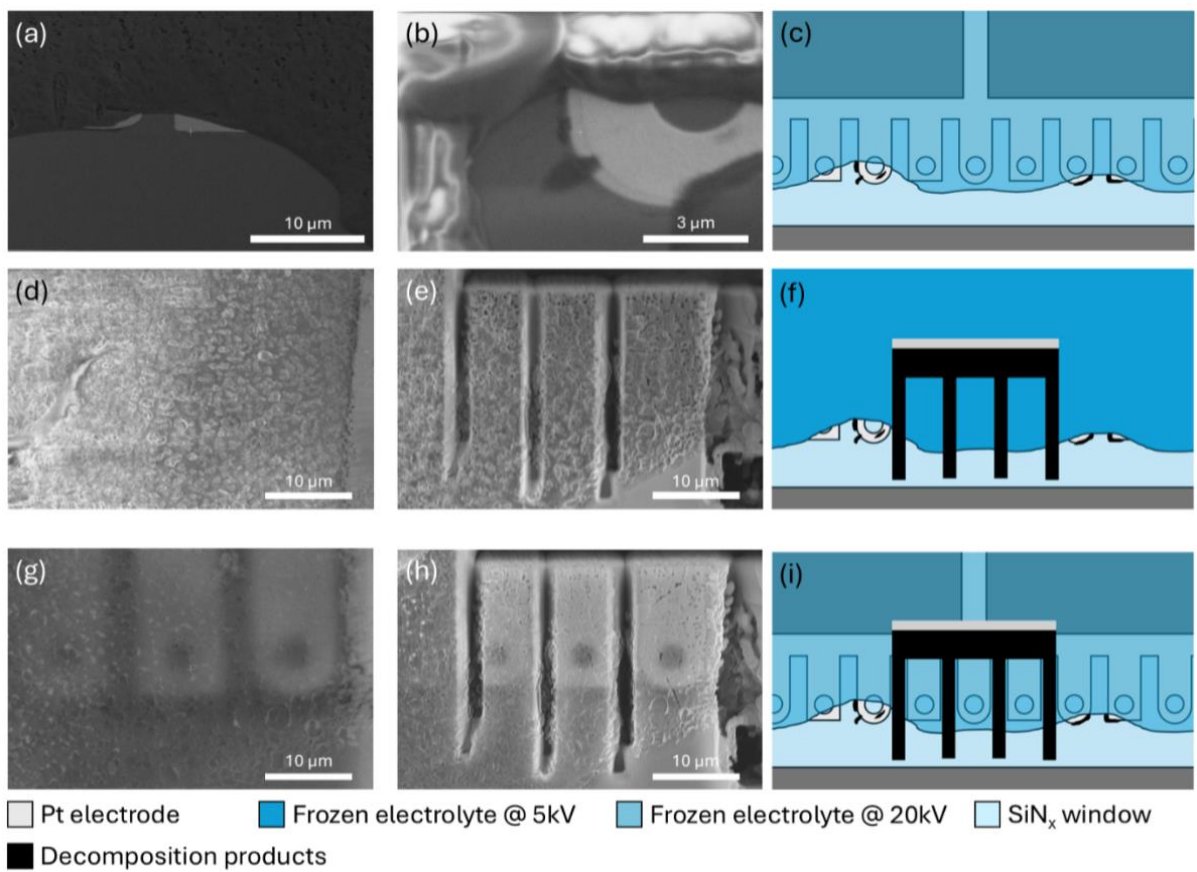

**Fig. S2:** The first part of the cryogenic FIB/SEM sample preparation procedure involving identifying the electrode-electrolyte interface and milling it free from the bulk sample is shown here. (a) and (b) show SEM micrographs of sections of the interface where the electrode is not covered by the thin layer of electrolyte. Cracking of the electrode and mossy Li deposits. This is represented schematically in (c) showing an overview of the ROI. (d) shows an SEM micrograph at 5kV of a region of the interface in which the electrodes are covered by the electrolyte. Due to the low accelerating voltage the electrodes cannot be seen under the electrolyte. In the same region the electrodes are subsequently milled free from the bulk using the ion beam (30 kV, 4 nA), and this can be seen in (e). This is illustrated in (f). SEM micrographs of the same region at 20 kV accelerating voltage is shown in (g) and milled free in (h). This is again illustrated in (i).

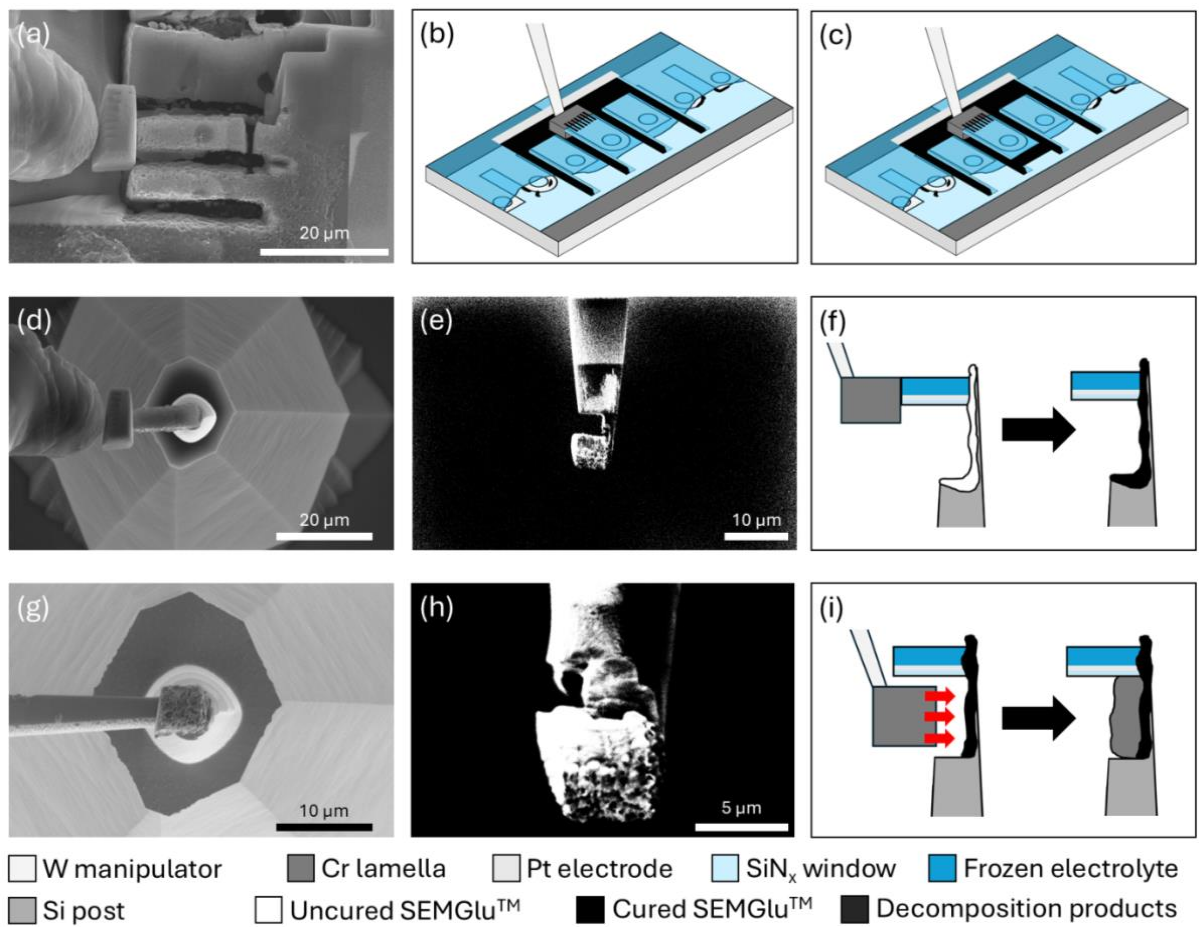

**Fig. S3:** The second part of the cryogenic FIB/SEM sample preparation procedure involving lifting out the identified electrode-electrolyte interface from the bulk, sticking it to a Si microarray post and filling in the underside using localised sputtering from a Cr lamella are shown here. (a) shows an SEM micrograph of an electrode covered in electrolyte being lifted out using redeposition welding (Xe, 30 kV, 30 pA) with a Cr lamella. This is represented schematically in (b) and (c), with the cr being attached to the sample in (b) and being milled free in (c). (d) shows an SEM micrograph of the lifted-out electrode being attached to the Si microarray post using SEMGlu$^{TM}$. The glue is cured using the Xe ion beam (30 kV, 30 pA, 30s). The sample is then subsequently milled free from the manipulator and remains in place with the correction orientation, as shown in the FIB micrograph in (e). This SEMGlu$^{TM}$ curing and attachment process is represented schematically in (f). The underside of the sample is subsequently filled in with Cr using localised sputtering from a Cr lamella. An SEM micrograph of the lamella underneath the sample can be seen in (g) and a FIB micrograph of the sample filled in with Cr is shown in (h). This fill in process is represented schematically in (i).

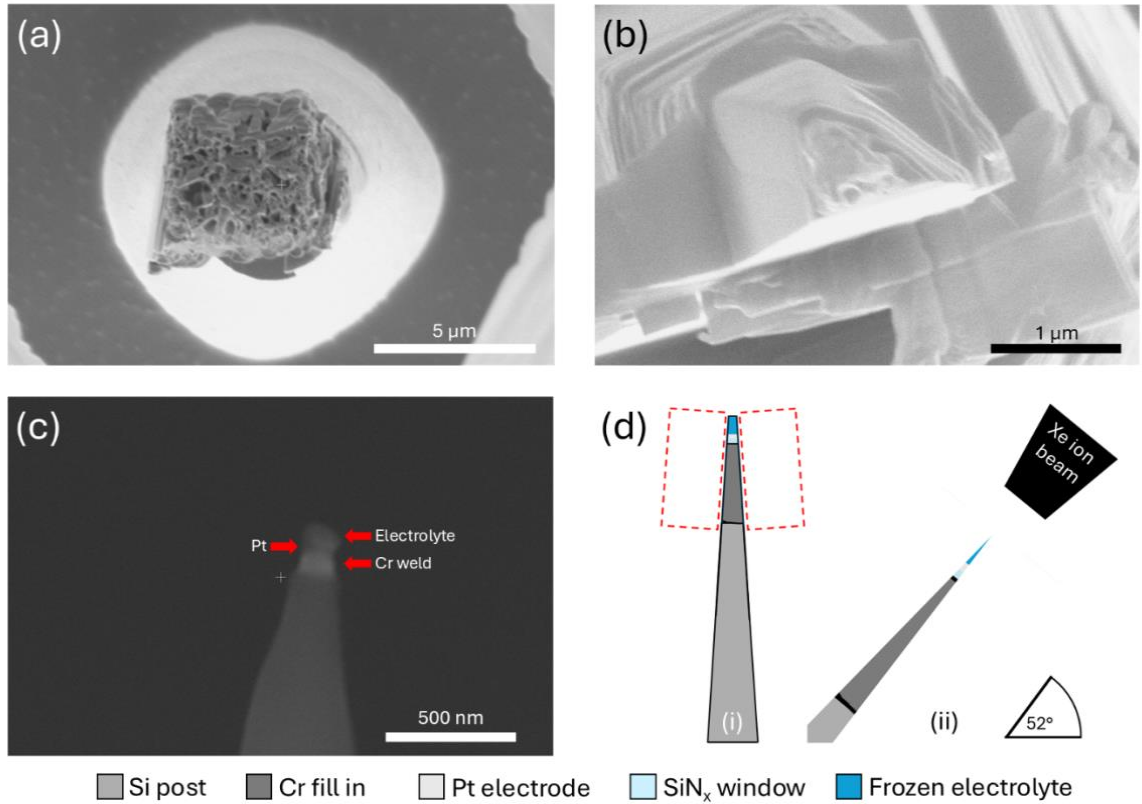

**Fig. S4:** The final part of the cryogenic FIB/SEM sample preparation procedure showing the preparation of the final needle sample. (a) shows an SEM micrograph of the sample milled to fit the shape of the post and cr weld. (b) is an SEM micrograph showing an example of the initial rough milling procedure at 0° stage tilt. Rectangular box milling patterns at shallower and shallower shape angles with period rotating was used to create the rough needle shape using the Xe ion beam (30 kV, 30pA -1nA), avoiding the apex to protect beam sensitive Li species on top of the electrode. Following rough milling the sample was tilted to 52° and fine milling using the ion beam (30kV, 10-30pA) was used to create the final needle shape. (c) shows an SEM micrograph of the final created needle with visible Cr weld, Pt electrode and electrolyte. The (i) rough milling and (ii) fine milling procedures are schematically shown in (d).

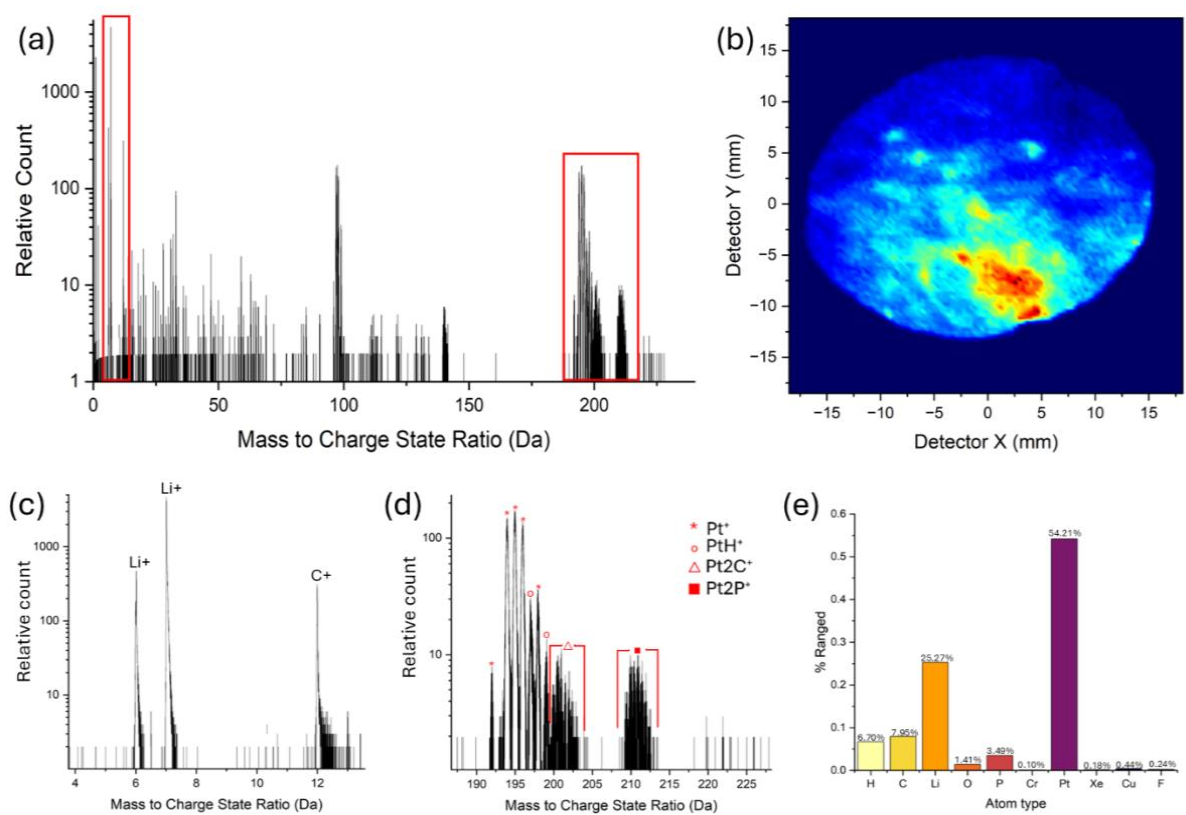

**Fig. S5:** (a) the produced mass spectrum and (b) detection hit map from the AP analysis. Two regions in (a) are highlighted in red and shown in (c) and (d) respectively. (c) shows relatively large amounts of elemental Li and C have been detected, while (d) highlights various detected Pt species including $Pt^+$, $PtH^+$, $Pt2C^+$ and $Pt2P^+$. (e) displays a bar chart plotting the percentage of ranged decomposed species against atom type, highlighting the total content of species in the generated mass spectrum.

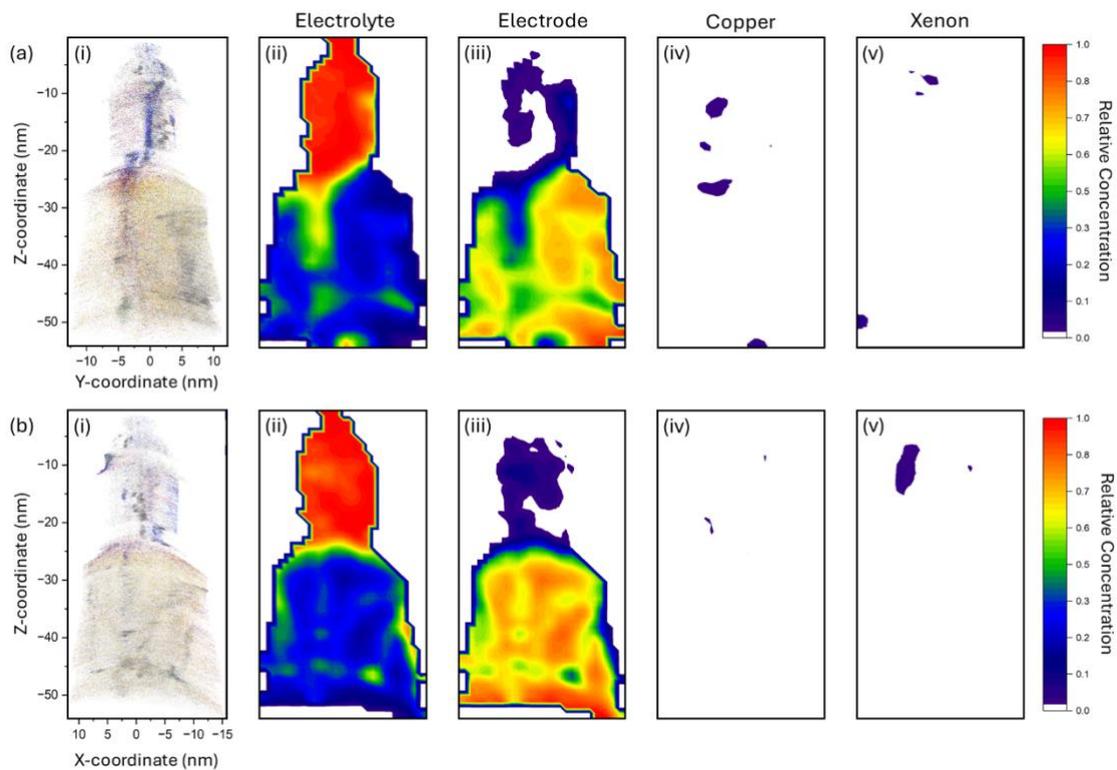

**Fig. S6:** Decomposed 2D contour plots in (a)Y-Z and (b)-X-Z orientations showing the spatial distribution in terms of relative concentrations of (ii) Electrolyte species (H,O,P,Li,F,P), (iii) Pt species, (iv) Copper and (v) Xenon. Relative concentrations are described in terms of atomic % per unit area.

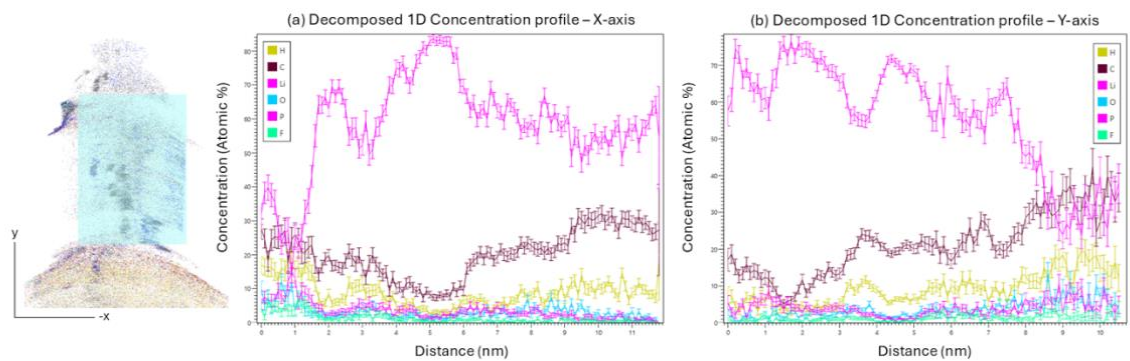

**Fig. S7:** Decomposed 1D concentration profiles in both (a) X and (b) Y directions plotting concentration in atomic % vs distance in nanometres. The ROI selected within the reconstruction for this analysis is shown by the blue box on the right. Within the plots, Pink corresponds to decomposed Li, bright green F, Dark pink P, Oxygen blue, Carbon dark purple and H light green.

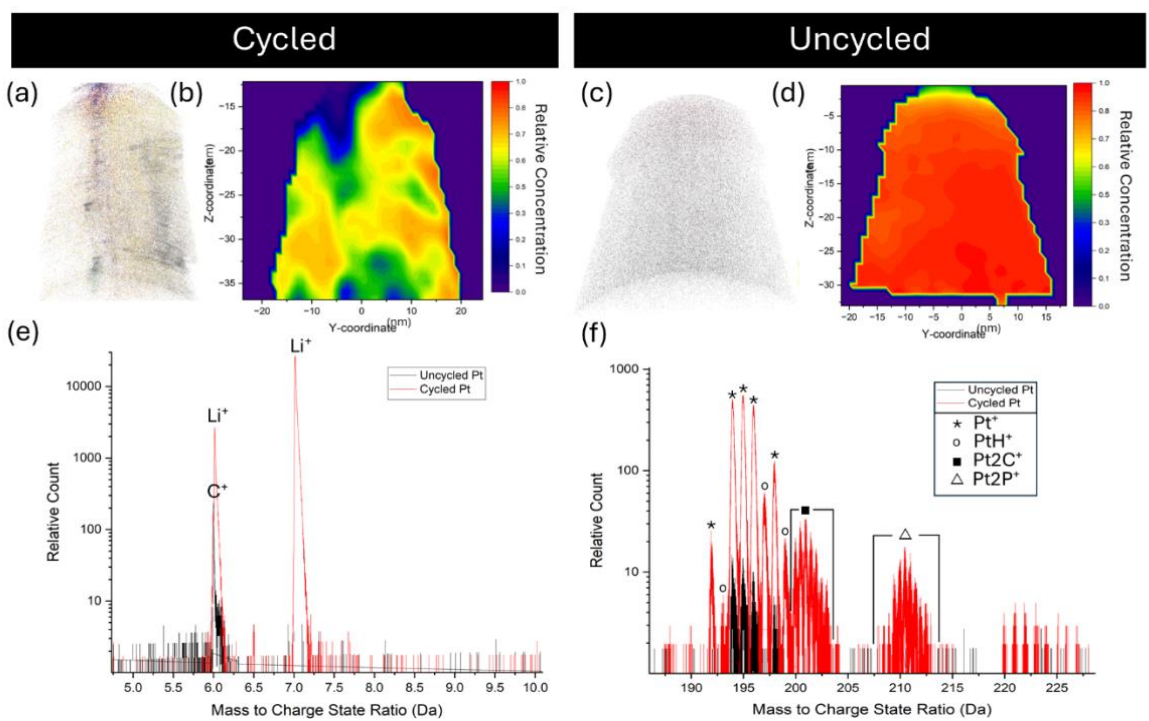

**Fig. S8:** (a) A 3D reconstruction of cycled Pt electrode and (b) a 2D contour plot detailing the relative concentrations of Pt containing species within the reconstruction. (c) A 3D reconstruction of an uncycled Pt electrode and (d) a 2D contour plot detailing the relative concentrations of Pt containing species within the uncycled electrode. (e) and (f) show sections of the overlayed mass spectra of both cycled (red) and uncycled (black) electrodes, highlighting the differences in composition. (e) focuses on the large amount of Li detected in the cycled electrode versus none in the uncycled and (f) shows the different Pt species detected including $Pt^+$, $PtH^+$, $Pt_2C^+$, and $Pt_2P^+$ in the cycled and just $Pt^+$ in the uncycled.

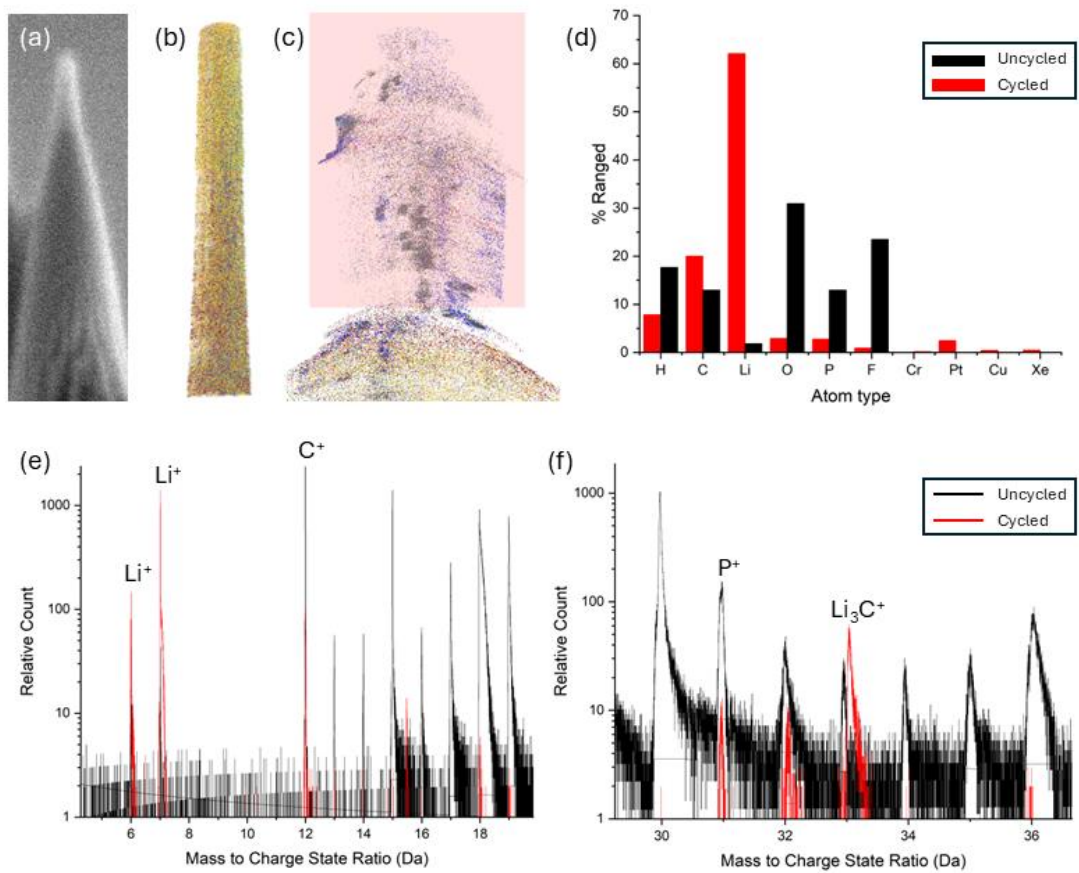

**Fig. S9:** A comparison between the *electrolyte* region of the sample and a needle of uncycled electrolyte. (a) shows an SEM micrograph of the needle of uncycled electrolyte created at cryogenic temperatures and the corresponding 3D reconstruction from APT analysis in (b). (c) shows the region of the sample the electrolyte needle is being compared to. (d) shows a bar chart plotting the percentage of ranged decomposed species against atom type in both needles, with the cycled sample in red and the uncycled in black. (e) and (f) show two select regions of the mass spectrums overlayed on one another comparing detected species, notably Li, C, P and $Li_3C$, with the cycled in red and uncycled in black.